\documentclass[preprint]{elsarticle}

\usepackage{graphicx}
\usepackage{pifont}
\usepackage{geometry}

\newcommand{\EDI}{School of Physics \& Astronomy, The University of Edinburgh, Edinburgh EH9 3JZ, UK}
\newcommand{\FREI}{Department of Physics, University of Freiburg, Hermann-Herder-Str. 3a, D-79104 Freiburg, Germany}
\newcommand{\LAPTH}{LAPTH, 9 Chemin de Bellevue BP 110, 74941 Annecy le Vieux, France}
\newcommand{\Nikhef}{Nikhef, H225 Science Park 105, 1098 XG Amsterdam, The Netherlands}
\newcommand{\unizh}{Institute for Theoretical Physics, Winterthurerstrasse 190, 8057 Zuerich, Switzerland}
\newcommand{\bea}{\begin{eqnarray}}
\newcommand{\eea}{\end{eqnarray}}
\newcommand{\nn}{\nonumber}

\begin{document}

\begin{frontmatter}

\title{Next-to-leading order QCD corrections to $pp\to b\bar{b} b\bar{b}+X$ \\ at the LHC: the quark induced case}

\author{T. Binoth}
\ead{thomas.binoth@ed.ac.uk}
\address{\EDI}
\author{N.~Greiner}
\ead{greiner@physik.unizh.ch}
\address{\unizh}
\author{A. Guffanti, J.~Reuter}   
\ead{alberto.guffanti@physik.uni-freiburg.de, reuter@physik.uni-freiburg.de}
\address{\FREI}
\author{J.-Ph. Guillet}
\ead{guillet@lapp.in2p3.fr}
\address{\LAPTH}
\author{T.~Reiter}
\ead{thomasr@nikhef.nl}
\address{\Nikhef}

\begin{abstract}
The production of two $b\bar{b}$ pairs is a prominent background for 
Higgs and New Physics searches in various extensions of the Standard Model. 
We present here the next-to-leading order QCD  corrections to the quark induced 
subprocess using the {\tt GOLEM} approach for the virtual corrections.
We show that our result considerably improves the prediction
and conclude that the inclusion of next-to-leading order effects is 
indispensable for reliable studies of $b\bar{b}b\bar{b}$ observables 
in hadronic collisions. 
\end{abstract}


\end{frontmatter}

\begin{flushright}
Edinburgh 09/16, FREIBURG-PHENO-09-05, LAPTH-1358/09, Nikhef 2009-026, ZU-TH 16/09
\end{flushright}


\section{Introduction}

With the LHC data taking on the horizon, the quest for precise predictions for
multi-parton processes becomes more and more pressing.
The confirmation of the Higgs sector of the Standard Model (SM) and the exploration 
of the TeV scale with respect to New Physics will ultimately depend on a 
very detailed understanding of signal and background processes 
\cite{:2008uu,Bern:2008ef,Buttar:2006zd,Campbell:2006wx}.
Especially in hadronic collisions, scale uncertainties lead to a strong logarithmic
dependence  if only leading order (LO) perturbation theory is applied. 
Whereas the inclusion of next-to-leading order (NLO) QCD effects is relatively 
straightforward for simple kinematic situations already for four parton
final states the number of results is very limited. 
The complexity of next-to-leading order computations with many external partons are many-fold.
It is not only necessary to provide a numerically stable implementation of the
ultra-violet (UV) renormalised one-loop amplitude, one also has to combine the different
parts of the computation such that infrared (IR) divergences are compensated
between virtual and real emission corrections. 
After IR cancellations the integrands are finite and
a stable integration over phase space can be achieved. In all respects
a lot of progress has happened in the last few years which point towards automated approaches
concerning the one-loop amplitudes 
\cite{Binoth:2009fk,Berger:2008sj,Bern:2007dw,Binoth:2005ff,Britto:2008vq,Denner:2005nn,Diakonidis:2008ij,Giele:2008bc,Hahn:2009bf,
Lazopoulos:2008ex,Ossola:2006us,Reiter:2009kb,Winter:2009kd} and IR subtraction modules \cite{Gleisberg:2007md,Seymour:2008mu,Frederix:2008hu,Hasegawa:2009zz,Czakon:2009ss,Frederix:2009yq}
which can be combined with publicly available Monte-Carlo tools 
\cite{Stelzer:1994ta,Maltoni:2002qb,Cafarella:2007pc,Kilian:2007gr,Sjostrand:2006za,
Bahr:2008pv,Gleisberg:2008ta}. 

Recent achievements
are the evaluation of the order $\alpha_s$ corrections to $pp\to b\bar{b}t\bar{t}$ 
\cite{Bredenstein:2008zb,Bredenstein:2009aj,Bevilacqua:2009zn} and $pp\to Wjjj$ \cite{Berger:2009ep,Berger:2009zg,KeithEllis:2009bu,Ellis:2009zw}. 
In both cases the inclusion of higher order effects  leads to the expected improvement 
in predictability of corresponding production rates which is of importance
for Higgs and New Physics searches.

In this paper we present a result of similar complexity which is relevant
in the context of Higgs searches in the Minimal Supersymmetric Standard Model (MSSM)
and two Higgs doublet extensions of the SM, where five scalar Higgs bosons are present.   
To scan and explore the parameter space in these models Higgs pair production is a 
relevant handle. As light Higgs bosons prefer to
decay into $b$-quarks in large parts of parameter space, experimental studies are so 
far strongly affected by the uncertainty of the Standard Model backgrounds, especially
$pp \to b\bar{b}b\bar{b}$, see for
example Ref.~\cite{Lafaye:2000ec}. Although mainly motivated by supersymmetry,
the four-$b$ final state also allows
the study of other interesting  beyond Standard Model (BSM) scenarios such
as hidden valley models, where decays of hadrons of an additional
confining gauge group can produce high multiplicities
of $b\bar{b}$ pairs~\cite{Bern:2008ef,Krolikowski:2008qa}.
This is the reason why the $pp\to b\bar{b}b\bar{b}$ process was added to the experimenters 
wish-list of relevant next-to-leading order computations \cite{Bern:2008ef}.

For the evaluation of the process $pp\to b\bar{b}b\bar{b}$ 
two partonic initial states contribute at tree level: $gg\to b\bar{b}b\bar{b}$ and $q\bar{q}\to b\bar{b}b\bar{b}$.
In this article we present the next-to-leading order QCD results for the quark induced process.   
Our calculation is based on the {\tt GOLEM} approach for one-loop amplitudes \cite{Binoth:2005ff,Binoth:2008gx,Binoth:1999sp,Binoth:2006hk} which
has already been applied in amplitude computations of various complexity \cite{Binoth:2005ua,Binoth:2006ym,Andersen:2007mp,Binoth:2007ca}.
The IR divergences are treated using  the Catani-Seymour dipole subtraction 
method \cite{Catani:1996vz} implemented 
in {\tt MadDipole} \cite{Frederix:2008hu}. For the evaluation of tree amplitude contributions
the matrix element generators {\tt Madgraph/MadEvent} \cite{Stelzer:1994ta,Maltoni:2002qb} and 
{\tt Whizard} \cite{Kilian:2007gr,Moretti:2001zz} are used.  
In the latter the relevant dipole subtraction terms have been included, too. 

In the next section we will briefly sketch how our result is obtained
and in section 3 numerical results are presented. Section 4 resumes the
paper. 

\section{Calculation}

\begin{figure}[ht]
\mbox{} \hspace{2.5cm}
\includegraphics[height=3.2cm]{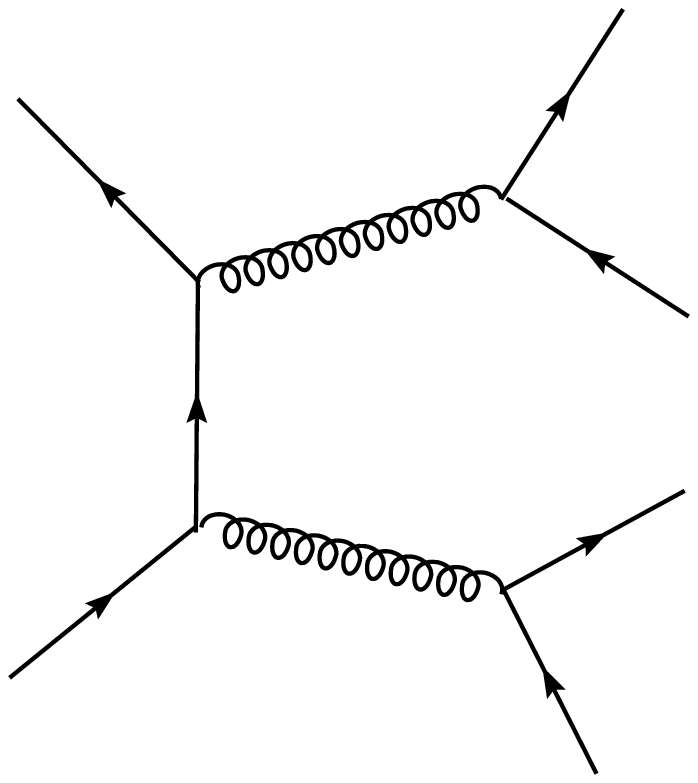} \hspace{2.5cm}
\includegraphics[height=3.2cm]{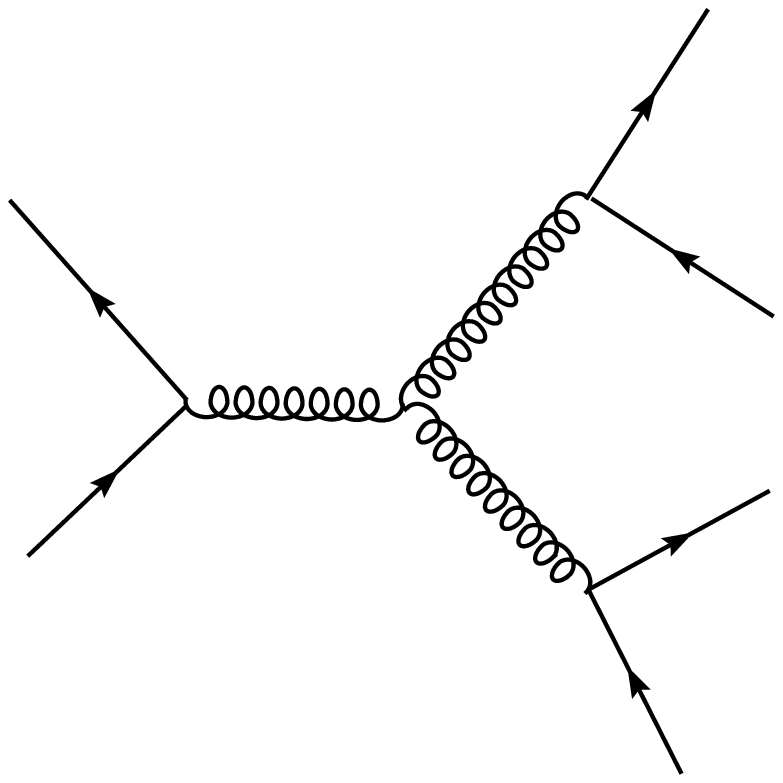}
\caption{Six-quark topologies which contribute to the leading order amplitude 
$q\bar{q}\to b\bar{b}b\bar{b}$.}
\label{fig:lo_topos}
\end{figure}

At leading order two topologies contribute to the amplitude $q\bar{q}\to b\bar{b}b\bar{b}$, 
see Fig. (\ref{fig:lo_topos}). 
As the parton distribution
functions for the $b$-quarks are very small we only consider the cases $q=u,d,s,c$.
We treat the $b$-quarks as massless which is a very good approximation for LHC kinematics 
and cuts as long as the $b$-quarks can be detected and are sufficiently separated in phase 
space. Effects of the heavy top quark are neglected altogether as they are numerically
not important.

 
\subsection{Virtual corrections}

\begin{figure}[ht]
\mbox{} \hspace{2.5cm}
\includegraphics[height=3.2cm]{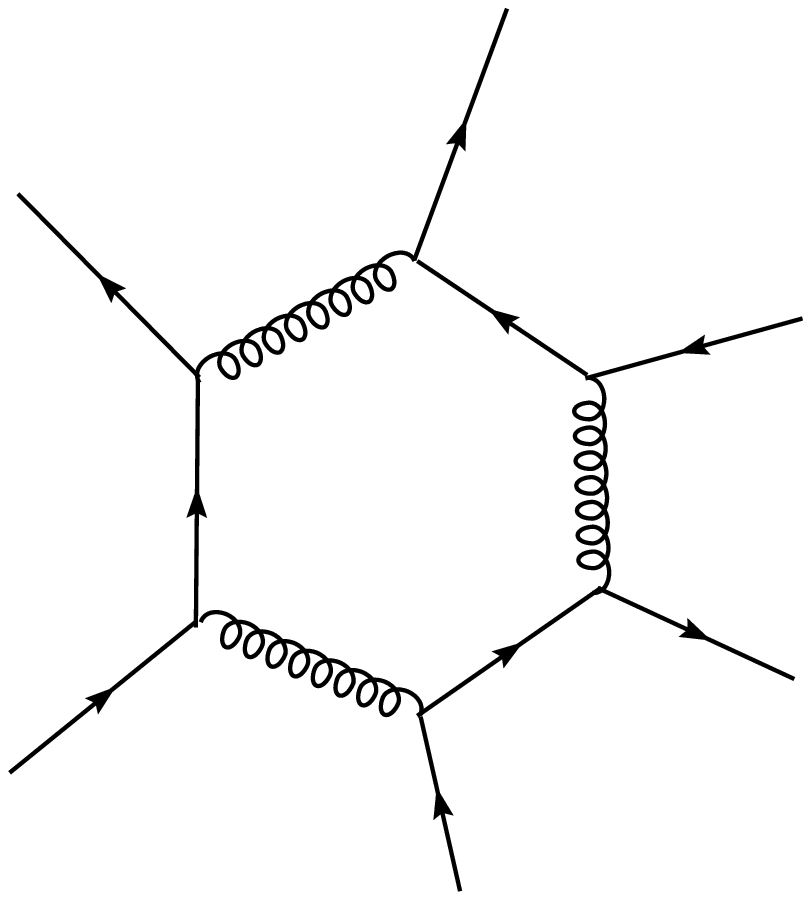} 
\hspace{2.5cm}
\includegraphics[height=3.2cm]{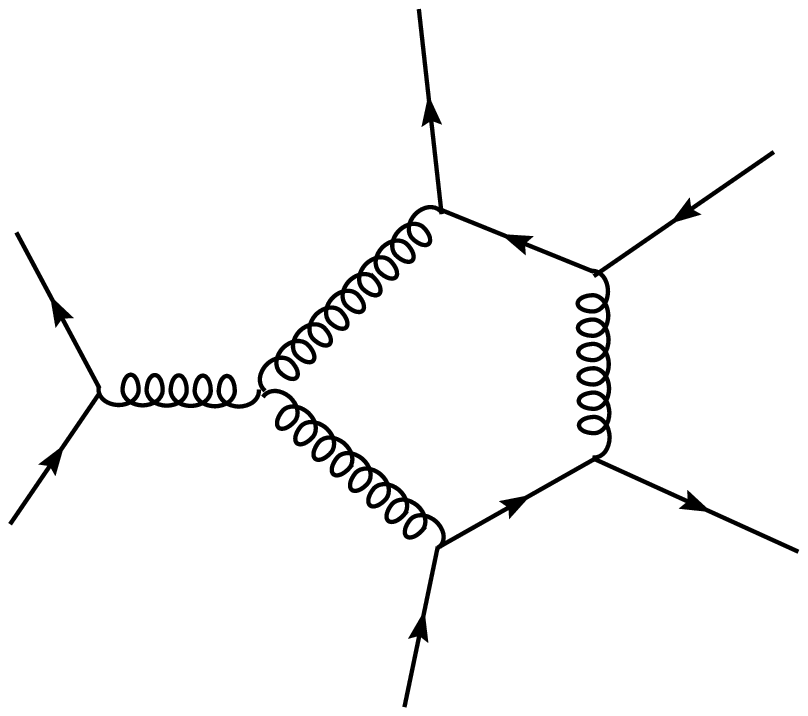}
\caption{Pentagon and hexagon one-loop topologies which contribute to the virtual corrections
to order ${\cal O}(\alpha_s^3)$.}
\label{fig:nlo_topos}
\end{figure}

The evaluation of the virtual corrections is the technically most challenging part
of the calculation. Our approach is based on the Feynman diagrammatic
representation of the amplitude. The most complicated topologies are 
of pentagon and hexagon  type as shown in Fig. (\ref{fig:nlo_topos}).
We evaluated all Feynman diagrams in two independent ways using the 
't~Hooft-Veltman scheme.
What is common to both methods is the projection onto different helicity and
colour structures.  
One strategy (1) is to convert the Feynman diagrams to a form factor
representation and to evaluate these and the resulting objects numerically. 
In this case we use Feynman diagram representations generated 
with {\tt QGRAF} \cite{Nogueira:1991ex}. {\tt FORM} \cite{Vermaseren:2000nd}
is used to map the helicity amplitudes to a tensor form factor
representation as defined in \cite{Binoth:2005ff,Reiter:2009kb,Reiter:2009dk}. The latter
is exported to {\tt Fortran95} code such that
it can be linked to the form factor library {\tt golem95} \cite{Binoth:2008uq}.
For the code generation we have developed a dedicated optimisation tool
which can also be used for other purposes \cite{Reiter:2009ts}.
The other strategy (2) is to perform
a symbolic tensor reduction down to genuine scalar integrals.
Here we use {\tt FeynArts} and {\tt FeynCalc} \cite{Hahn:1998yk} to generate the amplitudes.
The tensor integrals are reduced symbolically to
scalar integrals also using the formalism described in \cite{Binoth:2005ff}. 
As a scalar integral basis we  choose two- and three-point functions in $D=4-2\epsilon$-dimensions
and four-point functions in $6$ dimensions. In this way the extraction of 
infrared and ultraviolet poles is very transparent.  
In both strategies the reduction of the five-point tensor integrals    
is done without introducing inverse Gram-determinants.  

Without UV counterterms and IR subtraction terms
one has non-vanishing coefficients for double and single poles in $1/\epsilon$.
Adding in the UV counterterm 
and the Catani-Seymour IR insertion operator, $I(\epsilon)$ \cite{Catani:1996vz}, results
in the numerical cancellation of the pole part. This by itself is 
a stringent check on the algebra.

All Feynman diagrams are thus evaluated in two completely independent 
manners. Together with the cancellation of UV/IR poles and the test of symmetry
properties of helicity amplitudes the correctness of
the virtual amplitudes is well checked.

\subsection{Real emission corrections}

Apart from the one-loop corrections we have to consider  
processes with an extra parton in the final state. 
In this paper we restrict ourselves to $q\bar{q}$ initial states.
Further we are only  interested in observables with four well separated b-jets
where the $b$-quarks can be tagged.  The relevant amplitude is thus 
$q\bar{q} \to b\bar{b}b \bar{b}g$.
If the extra gluon is soft and/or collinear infra-red
divergences are treated by using the dipole subtraction  method of 
Catani-Seymour \cite{Catani:1996vz}.  
In our case 30 dipole subtraction terms are needed. In processes with so many dipoles the 
efficiency of the integration over phase space can be improved considerably
if the  dipole terms are only subtracted in restricted regions.
In the approach of \cite{Nagy:2005gn} a slicing parameter $\alpha$ controls the size 
of phase space where the subtraction terms are included.
This method is adopted here. Eventually the results must not depend on the slicing 
parameter $\alpha$. We have carefully checked that this is indeed the case.

For the evaluation of all tree like contributions of our process, i.e. the LO contribution
and the real emission corrections including dipole subtraction, we use adapted versions
of the tree level event generators {\tt Madgraph/MadEvent} \cite{Stelzer:1994ta,Maltoni:2002qb} and 
{\tt Whizard} \cite{Kilian:2007gr,Moretti:2001zz}. 
For the IR subtractions we use {\tt MadDipole} \cite{Frederix:2008hu} supplemented 
by integrated dipole contributions. Again we have two fully independent implementations
which should guarantee the correctness of the result.

\section{Results}

In this section we provide our prediction for the quark induced four $b$-quark 
production rate at next-to-leading order QCD for the LHC. We use $\sqrt{s}=14$ TeV
as center-of-mass energy for all results. To define a $b\bar{b}b\bar{b}$ event we use the 
following strategy.
\begin{itemize}
\item[1.)] First we apply a jet algorithm to identify the phase space region
where the $b$-quarks can be separated from the extra gluon jet by applying the $k_T$
jet algorithm as explained in \cite{Blazey:2000qt}. If a gluon is merged with a $b$-quark we
obtain an effective $b$-quark with a momentum $\tilde{p}_b = p_g + p_b$. 
\item[2.)] We apply separation cuts on the four resolved $b$-quarks which are defined by 
the transverse momenta, rapidities and separation parameters as
\bea
p_{T}(b_j) &>& 30 \;\; \textrm{GeV} \nn \\
|\eta(b_j)| &<& 2.5 \nn\\
\Delta R(b_i, b_j) &>& 0.8  \quad .
\eea 
\end{itemize}
The separation cut is defined in terms of the azimuthal and rapidity differences
of the (effectively) observed $b$-quarks, i.e. 
$\Delta R(b_i, b_j) = \sqrt{(\phi_i-\phi_j)^2 + (\eta_i-\eta_j)^2}$. 
Note that in the case of an unresolved gluon/$b$ system 
the effective momentum $\tilde{p}_b$ is used to test the cut conditions.  
We use the CTEQ6M parton distribution functions \cite{Pumplin:2002vw} 
with 2-loop running for $\alpha_s$ for both the LO and NLO order cross section evaluations.

\begin{figure}[ht]
\begin{center}
\includegraphics[height=7.0cm, clip=true]{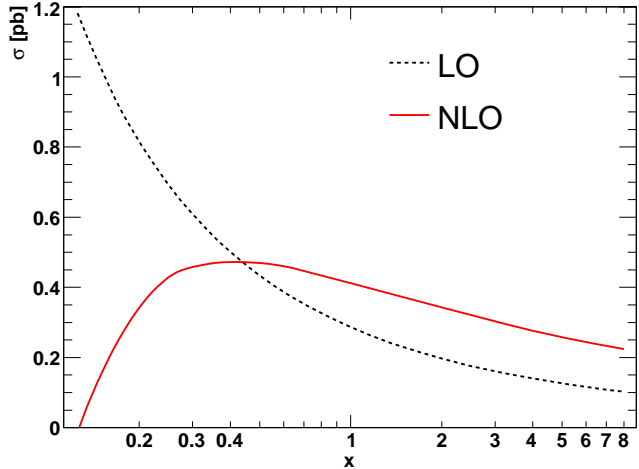}
\end{center}
\caption{The dependence of the LO and NLO prediction of $pp(q\bar{q})\to b\bar{b}b\bar{b} + X$ at the LHC
($\sqrt{s}=14$ TeV) on the renormalisation scale $\mu_R=x \mu_0$ 
with $\mu_0=\sqrt{\sum_j p_T^2(b_j)}$. The factorisation scale is fixed to $\mu_F=100$ GeV.}
\label{fig:scale_var}
\end{figure}

In Fig.~\ref{fig:scale_var} we show the renormalisation scale ($\mu_R$) dependence of 
our LO and NLO results. 
We vary the renormalisation scale around a standard scale $\mu_0$
which we define as the sum of the squared transverse $b$-quark momenta, $\mu_0 = \sqrt{\sum_j p_T^2(b_j)}$.
The factorisation scale is fixed to $\mu_F = 100$ GeV. As not all initial states are included
yet we postpone the study of the $\mu_F$ dependence to a forthcoming paper.
As expected the LO result shows 
a strong monotonous dependence on $x=\mu_R/\mu_0$.
Inclusion of the NLO corrections leads to a cancellation of the leading logarithmic
dependence such that the NLO result has now a plateau for a value around
$x \sim 1/2$, where the prediction shows the desired stability
under scale variation.

As a reference we provide cross section values for four scale choices
in Tab.~\ref{lonlo_numbers}.
\begin{table}[h]
\begin{center}
\begin{tabular}{|c|c|c|}
\hline
$\mu_R$ & $\sigma_{LO}$ [fb] & $\sigma_{NLO}$ [fb] \\
\hline
$2\mu_0$ & 197.1 &  342.7  \\
\hline
$\mu_0$ & 286.8 &  413.8  \\
\hline
$\mu_0/2$ & 434.0 &  463.0 \\
\hline
$\mu_0/4$ & 692.4 &  434.0 \\
\hline
\end{tabular}
\end{center}
\caption{Total LO and NLO cross sections for different renormalisation scale choices.
As a standard scale we use $\mu_0 = \sqrt{\sum_j p_T^2(b_j)}$. 
The factorisation scale is set to $\mu_F=100$ GeV. The precise definition of the
four-$b$ observable is given in the text. The error on all these numbers
is estimated to be well below 1\%.}
\label{lonlo_numbers}
\end{table}

\begin{figure}
\begin{center}
\includegraphics[height=7.cm, clip=true]{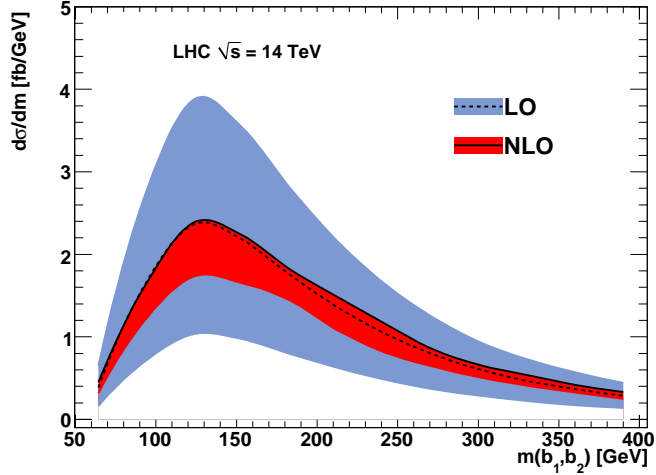}
\end{center}
\caption{Invariant mass ($m_{bb}$) distribution of the two 
leading $b$-quarks (see text). The LO/NLO bands are obtained by varying
the renormalisation scale $\mu_R$ between $\mu_0/4$ and $2\mu_0$ with 
$\mu_0 = \sqrt{\sum_j p_T^2(b_j)}$. The full (dashed) line shows the NLO 
(LO) prediction for the value $\mu_R=\mu_0/2$.}
\label{fig:mbb}
\end{figure}

Finally, we compare the LO and NLO prediction of the invariant mass distribution of 
the leading $b$-pair in Fig~\ref{fig:mbb}.
This invariant mass distribution is defined by the leading and sub-leading 
$b$ ($\bar{b}$) quark, i.e. we choose the two quarks with the largest $p_T$
and evaluate the invariant mass of this system.
We define an uncertainty band by varying the renormalisation scale $\mu_R$ 
around the scale between $\mu_0/4$ and $2\mu_0$, where again
$\mu_0 = \sqrt{\sum_j p_T^2(b_j)}$. This asymmetric scale choice
is motivated by the asymmetric shape of the NLO $\mu_R$ dependence, 
see Fig.~\ref{fig:scale_var}.
For the scale $\mu_R=\mu_0/2$ the
LO and NLO distributions look very similar which 
is indicated by the full (NLO) and dashed (LO) line
in Fig.~\ref{fig:mbb}. This is consistent
with Fig.~\ref{fig:scale_var} as for this value the
inclusive cross sections almost coincide.
The importance of the higher order terms is evident from
the uncertainty band which 
is considerably reduced after inclusion of the NLO QCD corrections. 

\section{Conclusion}

In this letter we have presented 
the  next-to-leading order QCD prediction for the quark induced  
four-$b$ cross section at the LHC. 
It is the first part of a complete
NLO QCD evaluation of $pp \to b\bar{b} b \bar{b} + X$. This final state 
is an important background for BSM Higgs searches at the LHC as  it is 
a prominent final state for Higgs pair production in large regions of parameter space.

From the technical point of view 
we have provided another example for the Feynman diagrammatic
approach being well capable to compute one-loop processes of
such a complexity. We have used the {\tt GOLEM} method implemented
in a highly automated framework such that further results 
can be obtained with the same set-up. Two independent evaluation
strategies guarantee the validity of the result.
For the tree-level amplitudes we use {\tt Madgraph/MadEvent} in combination 
with {\tt MadDipole} and {\tt Whizard}. For the latter the relevant
dipole terms have been added. 
We observe that the evaluation time for any observable
is dominated by the real emission part of the calculation.
Future improvements can be expected by phase-space parametrisations which 
are adapted to a given IR subtraction method. 
   
We have shown for the quark induced case that the
inclusion of the order $\alpha_s$ corrections leads to a significant
improvement of predictions for four $b$-observables at the LHC as
the renormalisation scale dependence is considerably reduced.
The same can be expected for the   the gluon induced case
which is under construction using the same methods.

\section*{Acknowledgements}

We would like to thank Thomas Gehrmann for interesting and useful discussions. 
T.B. thanks the University of Freiburg and Nikhef in Amsterdam, 
where parts of this work have been done, for their hospitality. 
T.B. is supported in parts by STFC, SUPA and the IPPP Durham. 
N.G. thanks the Universities Edinburgh, Freiburg and the IPPP in Durham
for supporting various visits where work on the project was done.
N.G.  was supported by the Swiss National Science Foundation
(SNF) under contract 200020-126691.
T.R. thanks the University of Zuerich and
J.R. wants to thank the Aspen and Les Houches Centers of Physics     
for their hospitality.
J.R. was partially supported by the Baden-Wuerttemberg Ministry of Science 
and Culture. Part of the computations were done on the ECDF cluster at the 
University of Edinburgh.

\end{document}